\documentstyle[aps,epsfig]{revtex}

\def\be{\begin{equation}}
\def\ee{\end{equation}}
\def\ba{\begin{eqnarray}}
\def\ea{\end{eqnarray}}
\def\ga{\mathrel{\raise.3ex\hbox{$>$\kern-.75em\lower1ex\hbox{$\sim$}}}}
\def\la{\mathrel{\raise.3ex\hbox{$<$\kern-.75em\lower1ex\hbox{$\sim$}}}}
\def\any{\mathrel{\raise.3ex\hbox{$<$\kern-.75em\lower1ex\hbox{$>$}}}}
\newcommand{\sect}[1]{\section{#1}\setcounter{equation}{0}}

\begin{document}
\begin{titlepage}
\rightline{CERN-TH/2001-304}
\rightline{hep-th/0110298}
\rightline{October 2001}
\begin{center}
 
\vspace{1cm}
 
\large {\bf Quest for Localized \bf{4-D} Black Holes
 in Brane Worlds}\\
\vspace*{5mm}
\normalsize

{\bf  Panagiota Kanti$^{1,2}$} and {\bf Kyriakos Tamvakis$^3$}

\smallskip
\medskip
$^1${\it CERN, Theory Division,\\
CH-1211 Geneva 23, Switzerland}

\vspace*{3mm}                      
                        
$^2${\it Scuola Normale Superiore, Piazza dei Cavalieri 7,\\
I-56126 Pisa, Italy}
 
\vspace*{3mm}                      
                        
$^3${\it Physics Department, University of Ioannina,\\
GR-451 10 Ioannina, Greece}

\smallskip
\end{center}
\vskip0.6in
 
\centerline{\large\bf Abstract}

We investigate the possibility of obtaining localized black hole solutions
in brane worlds by introducing a dependence of the four-dimensional
line--element on the extra dimension. An analysis, performed for the cases
of an empty bulk and of a bulk containing either a scalar or a gauge 
field, reveals that no conventional type of matter can support such a
dependence. Considering a particular ansatz for the five-dimensional
line--element that corresponds to a black hole solution with a 
``decaying'' horizon, we determine the bulk energy--momentum tensor
capable of sustaining such a behaviour. It turns out that an exotic,
shell-like distribution of matter is required. For such
solutions, the black hole singularity is indeed localized near
the brane and the spacetime is well defined near the AdS horizon,
in contrast to the behaviour found in black string type solutions.

\end{titlepage}

\sect{INTRODUCTION}

The recent studies on higher dimensional models of Gravitation, motivated by
the need  to explain the large difference in magnitude between the Planck
scale $M_P\sim 2\times 10^{18}$ GeV and the Electroweak scale of particle
physics, have resulted in a number of interesting theoretical ideas
\cite{large1,large2,RS1,RS2}. Among them, of particular interest is the
idea of an extra non-compact dimension \cite{RS2}. The models proposed
by Randall and Sundrum (RS) correspond to regions of $AdS$ space separated
by zero-thickness {\textit{$3$-branes}}. The Standard Model interactions are
confined on such a
3-brane, while gravitation, propagating in the five-dimensional
{\textit{bulk}}, is represented on the brane by an ordinary massless graviton
localized on it \cite{graviton}. The four-dimensional Planck mass is an
effective scale given in terms of a fundamental scale of five-dimensional
gravity and the AdS radius of the five-dimensional spacetime.  

The four-dimensional Schwarzschild metric describes the simplest case of
gravitational collapse in the standard four-dimensional world. In a
five-dimensional framework, as the Randall--Sundrum model, it would be natural
to ask if matter trapped on the brane, when it undergoes gravitational collapse,
could still be described by a Schwarzschild-type metric. The one-brane RS model
has  a factorized RS--Schwarzschild black hole solution\footnote{The
five-dimensional metric is
$g_{\mu\nu}=e^{-2\lambda |y|}g_{\mu\nu}^{(S)}\,\,,\,\,\,g_{55}=1$ with 
$g_{\mu\nu}^{(S)}={\rm diag} \bigl(-1+2M/r,\,(1-2M/r)^{-1},\,r^2,\,r^2
\sin^2\theta\bigr)$ being the four-dimensional Schwarzschild metric and
$\lambda$ the five-dimensional AdS radius.} corresponding to a
{\textit{black string}} infinite in the fifth dimension~\cite{CHR}. The
induced metric on the brane is purely Schwarzschild, respecting all the usual
astrophysical constraints~\cite{Giannakis}. Although the Ricci scalar and the
square of the Ricci tensor corresponding to this solution are everywhere finite,
the square of the Riemann tensor diverges at the AdS horizon at infinity as 
\be
R_{MNRS}R^{MNRS}\propto \frac{48 M^2 e^{4\lambda |y|}}{r^6}\,.
\label{riemann}
\ee
This singularity renders the above solution physically unsuitable and has led
to the speculation that there exists a localized {\textit{black cigar}}
solution which has a finite extension along the fifth dimension due to a 
Gregory-Laflamme type of instability~\cite{GL} near the AdS horizon.
The subject of black holes in the context of extra dimensions
has been extensively studied during the last years: black holes on
branes~\cite{bh-brane}, branes in a black hole background~\cite{branes-bh},
black hole solutions emerging from the superstring or supergravity
theory~\cite{super} and their stability~\cite{stability}, five- or
higher-dimensional black hole solutions~\cite{higher} and black holes
in scenarios with large extra dimensions~\cite{large} are some
of the topics studied so far.

The present article is devoted to a question that still remains open,
namely to the existence of black hole solutions that reduce to the standard
Schwarzschild black hole on the brane while having a finite extension along
the fifth dimension. This amounts to finding solutions with a
non-factorizable $y$-dependence of the four-dimensional
metric tensor $g_{\mu\nu}=e^{2A(y)}\hat{g}_{\mu\nu}(x,y)$, with
$\hat{g}_{\mu\nu}(x,0)=g_{\mu\nu}^{(S)}(x)$, in contrast to the  black
string case where $g_{\mu\nu}=e^{2A(y)}{g}_{\mu\nu}^{(S)}(x)$.
The assumed $y$-dependence of the four-dimensional metric tensor would
result in an additional scaling of the value of the black hole horizon
with respect to the extra dimension, independently of the one of the warp
factor. By choosing appropriately the $y$-dependence of
$\hat{g}_{\mu\nu}(x,y)$, this could lead to the vanishing of the horizon
away from the brane, but well before the AdS horizon, and thus to a possible
localization of the black hole along the extra dimension. The choice of
$\hat{g}_{\mu\nu}(x,y)$ should be consistent with the desired behaviour
of the metric tensor near or away from the brane as well as with the
boundary ({\it jump}) conditions that the presence of the 3-brane
introduces in the model.

Pursuing this idea, we first investigate the possibility that the
desired dependence of the four-dimensional metric tensor on the extra
dimension might be supported by an empty bulk or by a bulk containing
either a scalar or a gauge field. The ansatz for the metric tensor is
taken to be spherically symmetric, giving rise to $y$-dependent, neutral
or charged, generalizations of the Schwarzschild and AdS/dS--Schwarzschild
black hole solutions on the brane. Our analysis reveals that, for the
chosen class of $y$-dependence, an empty
bulk or a bulk dominated by a conventional form of matter cannot
support such a $y$-dependence. We, then, reverse the question and 
ask what kind of bulk matter can give rise to such $y$-dependent
solutions for the metric tensor. We consider a particular ansatz 
for the four-dimensional line--element that is characterized by a
rapidly ``decaying" value of the black hole horizon away from the
brane. The components of the corresponding bulk energy--momentum tensor
are determined and their behaviour, with respect to brane and bulk
coordinates, is examined. It turns out that the matter distribution
capable of sustaining localized black hole solutions is an exotic,
shell-like one, localized around the brane and vanishing asymptotically
at large distances from both the black hole horizon and the location
of the brane. This particular solution is free from the singularity
at the AdS horizon found in black string solutions and has the
black hole singularity localized near the brane. However, it is
accompanied by the appearance of a second, localized singularity at the
black hole horizon due to the singular behaviour of the bulk energy-momentum
tensor at the same point. We argue that this singularity is only due to
the particular choice of the five-dimensional spacetime and
that a more complicated choice, describing a four-dimensional black
hole solution on the brane without possessing a horizon, might not
have the same problem.

The paper has the following structure: in Sec. 2 we present the theoretical
framework of our model, as well as the ansatz for the five-dimensional
metric tensor, and derive the equations of motion. In Secs. 3, 4 and 5,
we look for $y$-dependent solutions for the four-dimensional metric tensor
in the case of an empty bulk and of a bulk containing either a scalar or
a gauge field, respectively. Section 6 contains the analysis of the
alternative approach of demanding a localized black hole solution
and examining the nature of the corresponding bulk matter. We present
our conclusions in Sec. 7.

\sect{EQUATIONS OF MOTION}

We shall consider the general framework defined by the five-dimensional
Einstein action in the presence of a $3$-brane located at an arbitrary point
chosen to be $y=0$, namely
\be
S=- \int d^4x\,dy\,\sqrt{-g}\,\biggl\{-\frac{R}{2 \kappa^2}
+ \Lambda_B + \frac{\sigma}{\sqrt{g_{55}}}\,\delta(y) - {\cal L}_B \biggr\}\,,
\label{action1}
\ee
where $\kappa^2=8\pi G_5$, with $G_5$ the five-dimensional Newton's
constant. In the above, $\Lambda_B$ stands for a {\textit{bulk cosmological
constant}}, while any additional existing {\textit{bulk matter}} is represented
by the Lagrangian  ${\cal L}_B$. The constant $\sigma$ represents the positive
{\textit{brane tension}}. The general form of the metric will be taken to be
\be
ds^2=e^{2A(y)}\hat{g}_{\mu\nu}(x,y)dx^{\mu}dx^{\nu}+dy^2\,. 
\ee
Throughout this article, we will be interested in solutions that are spherically
symmetric on the brane, we therefore move to the ansatz
\be
ds^2=e^{2 A(y)}\,\biggl\{-U(r,y)\,dt^2 +
\hat{U}(r,y)^{-1}\,dr^2 + r^2(d\theta^2 + 
\sin^2\theta\,d\varphi^2) \biggl\} + dy^2\,.
\label{metric}
\ee
In addition, we assume that the spacetime will remain invariant under a mirror
transformation $y \rightarrow -y$.
The case of the {\textit{black string}} solution~\cite{CHR} is recovered for
the Randall--Sundrum choice of the warp factor $A(y)=-\lambda |y|$, with
$\lambda^2=\kappa^2 |\Lambda_B|/6$, and 
\be
\hat{U}(r,y)=U(r,y)=1-\frac{2M}{r}\,.
\label{bstring}
\ee
Another particular type of solutions that arises as a special case of the above
metric ansatz is the RS--AdS/dS--Schwarzschild one \cite{Kaloper,Kim,KR,HK} for 
\be
e^{A(y)}= \cosh (\lambda y) - \frac{\kappa \sigma}{\sqrt{6 |\Lambda_B|}}
\,\sinh (\lambda |y|)
\label{AdS}
\ee
and
\be
\hat{U}(r,y)=U(r,y)=1-\frac{2M}{r} - \Lambda r^2\,,
\label{4DAdS}
\ee
where $\Lambda$ is the four-dimensional cosmological constant given by
\be
\Lambda = \frac{\kappa^2}{6}\,\Bigl(\frac{\kappa^2 \sigma^2}{6} -|\Lambda_B|
\Bigr)\,.
\ee
Similarly to the RS--Schwarzschild solutions, the above solutions are
characterized by finite $R$ and $R_{MN} R^{MN}$, while the square of the
Riemann tensor is given by Eq. (\ref{riemann}) where $e^{4\lambda |y|}$ is
replaced by $e^{-4 A(y)}$ according to Eq. (\ref{AdS}). The de Sitter
solution ($\Lambda>0$) is characterized by a true singularity at a finite 
distance $y=y_0$ from the brane, with $y_0$ defined by
\be
\tanh (\lambda |y_0|) = \sqrt{\frac{6 |\Lambda_B|}{\kappa^2 \sigma^2}}\,,
\ee
thus causing the square of the Riemann tensor to diverge and exhibiting the
same unnaturalness as the RS--Schwarzschild solution. For the Anti-de Sitter
solution ($\Lambda<0$), all gravitational scalars are well defined
everywhere; however, the divergence of the warp factor itself at infinite
distance from the brane necessitates the introduction of a second brane for
the recovery of the conventional four-dimensional gravity (for alternative
methods, see Refs.~\cite{kkop,KR}).

In the following analysis, we will be looking for $y$-dependent (neutral or
charged) generalizations of the above solutions that reduce to conventional
four-dimensional black hole solutions on the brane (Schwarzschild,
Reissner--Nordstrom or AdS/dS--Schwarzschild). In order to simplify the equations
of motion resulting from the above action (\ref{action1}) and the metric
ansatz (\ref{metric}), we will consider a restricted  case, namely $U=\hat{U}$.
Under this assumption, Einstein's equations of motion, written in terms of the
Einstein tensor $G_{MN}=R_{MN}-\frac{1}{2}g_{MN}R$ and a {\textit{bulk
energy--momentum tensor}}, are
\ba
&~& G^t_t=6A'^2 + 3A'' - \frac{2 A' U'}{U} + \frac{ 3 U'^2}{4U^2}
-\frac{U''}{2U} + \frac{U-1}{e^{2A}\,r^2} + \frac{1}{e^{2A}\,r}
\frac{\partial U}{\partial r} = \kappa^2 \,[-\Lambda_B-\sigma \delta (y)
+ T^{t}_t]
\label{tt} \\[4mm]
&~& G^r_r=6A'^2 + 3A'' + \frac{2 A' U'}{U} - \frac{U'^2}{4U^2}
+ \frac{U''}{2U} + \frac{U-1}{e^{2A}\,r^2} + \frac{1}{e^{2A}\,r}
\frac{\partial U}{\partial r} = \kappa^2 \,[-\Lambda_B-\sigma \delta(y)
+ T^{r}_r]
\label{rr} \\[4mm]
&~& G^\theta_\theta = 6A'^2 + 3A'' + \frac{U'^2}{4U^2}
+ \frac{1}{e^{2A}\,r}\,\frac{\partial U}{\partial r} + 
\frac{1}{2 e^{2A}}\,\frac{\partial^2 U}{\partial r^2} 
= \kappa^2 \,[-\Lambda_B-\sigma \delta(y) + T^{\theta}_\theta]
\label{th-th} \\[4mm]
&~& G^\varphi_\varphi= G^\theta_\theta =
\kappa^2 \,[-\Lambda_B -\sigma \delta(y) + T^{\varphi}_\varphi]
\label{ph-ph} \\[4mm]
&~& G^5_5=6A'^2 - \frac{U'^2}{4U^2} + \frac{U-1}{e^{2A}\,r^2} + 
\frac{2}{e^{2A}\,r} \frac{\partial U}{\partial r} + 
\frac{1}{2 e^{2A}}\,\frac{\partial^2 U}{\partial r^2}
= \kappa^2 \,[-\Lambda_B + T^{5}_5]
\label{55} \\[4mm]
&~& G_{r5}= -\frac{1}{U r} \frac{\partial U}{\partial y} -
\frac{1}{2U} \frac{\partial^2 U}{\partial r \,\partial y}=
\kappa^2 \,T_{r5} \label{r5}\,, 
\ea
where a prime denotes differentiation with respect to $y$ and  $T^{M}_N$
the bulk distribution of energy associated with the Lagrangian ${\cal L}_B$.

Taking the warp factor to be the Randall--Sundrum one, namely 
$e^{A(y)}=e^{-\lambda |y|}$, or the RS--AdS/dS--Schwarzschild one (\ref{AdS}),
the equations of motion reduce to a set of partial differential equations for
$U(r,y)$. In the former case, the terms involving solely derivatives of 
the function $A(y)$ cancel exactly the bulk cosmological constant term
while in the latter case, we need to make the substitutions
$T^\mu_\nu \rightarrow T^\mu_\nu -3 \Lambda e^{-2A}$ and 
$T^5_5 \rightarrow T^5_5 -6 \Lambda e^{-2A}$. The components of Einstein's
equations containing second derivatives with respect to the extra coordinate
will give rise to the {\it jump} conditions at the location of the brane. For
the RS--Schwarzschild solution, the fine-tuning $\lambda =\kappa^2 \sigma /6$
is still required, while for the RS--AdS/dS--Schwarzschild one it is the
location of the minimum/singularity, and thus the size of the extra dimension,
that is fixed in terms of the fundamental parameters of the theory. In both cases,
however, the same equations demand the {\it continuity} of the metric function
$U(r,y)$ along the extra dimension, a constraint that will prove to be
of great importance in the following analysis.


We shall now consider and analyse the equations for $U(r,y)$ in
various cases, namely, in the case of empty bulk and in the cases of bulk
containing either a scalar or a gauge field.

\sect{EMPTY BULK}

 In this case, all the components of
$T^M_N$ are assumed to be zero. The off-diagonal component 
(\ref{r5}) can then be easily integrated to give the solution
\be
U(r,y)=\frac{\lambda(y)}{r^2} + f(r)\,,
\label{solU}
\ee
where $\lambda(y)$ and $f(r)$ are arbitrary functions. We impose the
condition that, in the limit $y \rightarrow 0$, $\lambda(y)$ is a 
well-defined function. Moreover, in order to ensure the correct
asymptotic behaviour at infinity for each AdS slice, we 
rewrite the above solution as
\be
U(r,y)=1 + \frac{\lambda(y)}{r^2} + \sum_n \alpha_n\,r^n\,,
\label{ans1}
\ee
where $n$ is an integer number (strictly negative for the Schwarzschild and
Reissner--Nordstrom black hole solutions on the brane, both negative and
positive for the AdS/dS--Schwarzschild ones) and $\alpha_n$ arbitrary
constants. Taking the sum of Eqs. (\ref{th-th}) and (\ref{55}), we find the
constraint~\footnote{Whenever necessary, the upper entry of a two-line
matrix will stand for the result for a ``bent" 3-brane ($\Lambda \neq 0$)
while the lower entry will stand for a flat 3-brane ($\Lambda=0$).}
\be
\frac{U-1}{e^{2A}\,r^2} + 
\frac{3}{e^{2A}\,r} \frac{\partial U}{\partial r} + 
\frac{1}{e^{2A}}\,\frac{\partial^2 U}{\partial r^2}=
\biggl\{ \begin{tabular}{c} $-9 \Lambda e^{-2A}$ \\
$0$ \end{tabular} \biggr \}\,.
\label{empty}
\ee
Upon substituting the solution (\ref{ans1}) in the above expression,
it reduces to 
\be
\frac{\lambda(y)}{r^4} + \sum_n \alpha_n\,r^{n-2} (n+1)^2=
\biggl\{ \begin{tabular}{c} $-9 \Lambda$  \\
$0$  \end{tabular} \biggr \}
\ee
with the unique solution
\be
\lambda(y) =0 \,, \quad \quad \biggl\{ \begin{tabular}{c}
$n=-1,\,a_{-1}={\rm const.} \quad {\rm and} \quad n=2,\,a_2=-\Lambda $ \\[2mm]
$n=-1,\,a_{-1}={\rm const.}$ \end{tabular} \biggr \}\,,
\ee
which is nothing else than the RS--AdS/dS--Schwarzschild and the
{\textit{black string}} RS--Schwarzschild solution, respectively,
for $a_{-1}=-2M$. We therefore conclude that, for the assumed asymptotic
behaviour, any $y$-dependence, described by the ansatz (\ref{metric})
with $\hat g_{tt}=-1/\hat g_{rr}$, is ruled out in the case of an empty
bulk (see also \cite{GR}) and a non-trivial energy distribution is
required for this purpose.

\sect{A BULK SCALAR FIELD}

In the presence of a five-dimensional scalar field in the bulk, the components
of $T^{M}_N$ assume the form
\be
T^{t}_t  =  -\frac{U}{2}\,e^{-2A}
\,(\partial_r\phi)^2 - \frac{\phi'^2}{2}- V(\phi)\,, \qquad
T^{r}_r = \frac{U}{2}\,e^{-2A}
\,(\partial_r\phi)^2 - \frac{\phi'^2}{2} -V(\phi) 
\ee
\be
T^{\theta}_\theta =T^{\varphi}_\varphi=
T^{t}_t\,, \qquad 
T^{5}_5  = -\frac{U}{2}\,e^{-2A}
\,(\partial_r\phi)^2 + \frac{\phi'^2}{2} - V(\phi)\,, \qquad
T_{r5} = (\partial_r\phi)\,\phi'\,.
\ee

Let us first assume that the bulk scalar field depends only on the extra
coordinate $y$. Then, the off-diagonal component of $T^{M}_N$ is
again zero, which leads once again to the solution (\ref{ans1}) for the
metric function $U(r,y)$. Taking again the sum of Eqs. (\ref{th-th}) and
(\ref{55}), we now find 
\be
\frac{U-1}{e^{2A}\,r^2} + 
\frac{3}{e^{2A}\,r} \frac{\partial U}{\partial r} + 
\frac{1}{e^{2A}}\,\frac{\partial^2 U}{\partial r^2}= 
-2 \kappa^2 \,V(\phi) + \biggl\{ \begin{tabular}{c} $-9 \Lambda  e^{-2A}$
\\ $0$ \end{tabular} \biggr \}\,.
\ee
or
\be
\frac{\lambda(y)}{r^4} + \sum_n \alpha_n\,r^{n-2} (n+1)^2=
-2 \kappa^2 \,V(\phi) \,e^{2A} + \biggl\{ \begin{tabular}{c} $-9 \Lambda$
\\ $0$ \end{tabular} \biggr \}\,.
\ee
The right-hand side (r.h.s.) of the above equation is a pure function of $y$
while the left-hand side is not. Therefore, the only solution is 
\be
\lambda(y) = V(\phi)= 0\,, \quad \quad \biggl\{ \begin{tabular}{c}
$n=-1,\,a_{-1}={\rm const.} \quad {\rm and} \quad n=2,\,a_2=-\Lambda $ \\[2mm]
$n=-1,\,a_{-1}={\rm const.}$ \end{tabular} \biggr \}\,,
\ee
which reduces again to the RS--AdS/dS--Schwarzschild and RS--Schwarzschild solutions,
respectively, excluding any $y$-dependence of the 4D metric function.

The solution (\ref{ans1}) for $U(r,y)$ still holds if we assume instead that
the bulk scalar field depends only on the brane coordinate $r$. In that
case, we proceed as follows. Rearranging appropriately the pairs of
Eqs. (\ref{tt}, \ref{rr}) and (\ref{th-th}, \ref{55}), we find the
constraints
\be
\frac{2(U-1)}{e^{2A}\,r^2} + 
\frac{2}{e^{2A}\,r} \frac{\partial U}{\partial r} + 
\frac{U'^2}{2U^2}= -2 \kappa^2 \,V(\phi) + 
\biggl\{ \begin{tabular}{c} $-6 \Lambda  e^{-2A}$
\\ $0$ \end{tabular} \biggr \}\,, \qquad 
\frac{U'^2}{2U^2}=\frac{U-1}{e^{2A}\,r^2} + 
\frac{1}{e^{2A}\,r} \frac{\partial U}{\partial r} +
\biggl\{ \begin{tabular}{c} $3 \Lambda  e^{-2A}$
\\ $0$ \end{tabular} \biggr \}\,,
\label{con12}
\ee
respectively. Subtracting the above two constraints, we 
find 
\be
\frac{U-1}{r^2} + 
\frac{1}{r}\,\frac{\partial U}{\partial r} = 
-\frac{2}{3}\,e^{2A} \kappa^2 \,V(\phi) + \biggl\{ \begin{tabular}{c}
$-3 \Lambda$ \\ $0$ \end{tabular} \biggr \}\,,
\label{con-diffa}
\ee
which leads to
\be
-\frac{\lambda(y)}{r^4} - \sum_n \alpha_n\,r^{n-2} (n+1)=
-\frac{2}{3}\,e^{2A} \kappa^2 \,V(\phi) + 
\biggl\{ \begin{tabular}{c} $-3 \Lambda$
\\ $0$ \end{tabular} \biggr \}\,.
\ee
In this case, the r.h.s. of the above equation is a product of the
discontinuous warp factor times a function of $r$. Since, according to the
{\it jump} conditions, the l.h.s. is continuous in $y$, this equation
cannot be maintained with a non-trivial potential and a non-constant $\lambda$.
Thus, in this case also we have only the {\textit{black string}} solution
and its AdS/dS generalization (\ref{AdS})-(\ref{4DAdS}).

Next, we consider the more general case where the bulk scalar field depends
on both coordinates. The rearrangement of Eqs. (\ref{tt}, \ref{rr}) and
(\ref{th-th}, \ref{55}), leads again, in the same way as above, to the 
constraint (\ref{con-diffa}). The off-diagonal component of Einstein's
equations (\ref{r5}) cannot be integrated now due to the presence of the
non-vanishing component $T_{r5}$. However, motivated by the solution for
$U$ derived in earlier cases, we assume the form:
\be 
U(r,y)=1- \sum_n \alpha_n(y)\,r^n\,,
\label{ans2}
\ee
where $n$ is again an integer number. Taking the sum of Eqs. (\ref{th-th})
and (\ref{55}) and using the constraint (\ref{con-diffa}), we obtain 
\be
\kappa^2\,U\,(\partial_r \phi)^2= \frac{2(U-1)}{r^2}-
\frac{\partial^2 U}{\partial r^2}=
\sum_n \alpha_n(y)\,r^{n-2}\,(n+1)(n-2)\,,
\label{partial-phi}
\ee
which is indeed positive for $\alpha_n >0$ and $n<-1$ and/or $n>2$. Note that,
in agreement with the conclusion drawn in the case of an empty bulk, in the
absence of a bulk scalar field, only the terms $n=-1$ and $n=2$, the mass term
and the cosmological constant term, survive in the expression of the 4D metric
function. Subtracting finally Eqs. (\ref{tt}) and (\ref{rr}), we find
\be
\frac{4 A' U'}{U} -\frac{U'^2}{U^2} + \frac{U''}{U}=
\kappa^2\,U\,e^{-2A}\,(\partial_r \phi)^2
\ee
or, using Eq. (\ref{partial-phi}), 
\be
-\Bigl(1-\sum_m \alpha_m\,r^m\Bigr)\,\sum_n r^n (4 A' \alpha_n' + \alpha_n'')-
\sum_{n,m} \alpha_n' \alpha_m' r^{n+m}= e^{-2 A} \sum_n \alpha_n\,r^{n-2}
\,(n+1)\,(n-2)\,\Bigl(1-\sum_m \alpha_m\,r^m \Bigr)^2\,.
\ee
Collecting terms of order $r^n$, we are forced to impose the condition
$4 A' \alpha_n' + \alpha_n''=0$, which, apart from being contradictory to our
assumptions for continuous $\alpha_n(y)$, it is also  disastrous in two ways:
first, it leads to the result $\alpha_n(y) \sim e^{-4A}$, which has a
behaviour opposite to that of the warp factor --- namely, as gravity becomes
weaker, the ``charges" at infinity become larger; secondly, in that case,
what is left from the above equation can be rewritten as
\be
\kappa^2\,U\,e^{-2A}\,(\partial_r \phi)^2 = -\frac{U'^2}{U^2}\,,
\ee
which is inconsistent given the reality of the metric function and of the 
bulk scalar field.

\sect{A BULK GAUGE FIELD}

Next, we consider the case where,
in addition to the bulk cosmological constant, a  bulk gauge field gives
rise to a non-constant energy--momentum tensor
\be
T_{MN}=F_{MP}\,F^{\,\,\,\,P}_{N}-\frac 14\,g_{MN}\,F_{AB}\,F^{AB}\,.
\ee

We start our analysis with some simple, but characteristic, choices for
the gauge field configuration. We first assume that the only non-vanishing
component of the gauge field is $A_0=a(y)$, inspired by the choice made
by Visser in Ref.~\cite{visser}. For this particular choice, we
obtain the following components of the bulk energy--momentum tensor:
\be
T^{t}_t=T^{5}_5=- \frac{e^{-2A} a'^2}{2U}\,, \qquad 
T^{r}_r=T^{\theta}_\theta=T^{\varphi}_\varphi=
\frac{e^{-2A} a'^2}{2U}\,.
\ee
Since the field strength has only one non-vanishing component,
$F_{05}$, the off-diagonal energy--momentum tensor component $T_{r5}$ is
identically zero. The integration of Eq. (\ref{r5}), then, leads again
to the solution (\ref{ans1}) for the metric function $U(r,y)$. 
Despite the presence of a bulk $T_{MN}$, the analysis here closely
follows the one performed in the case of
an empty bulk. Since $T^{\theta}_\theta$ and $T^{5}_5$
are equal but of opposite sign, the sum of Eqs. (\ref{th-th}) and (\ref{55})
leads to the constraint (\ref{empty}) and thus to the unique
RS--AdS/dS--Schwarzschild and RS--Schwarzschild solutions.

An alternative choice could be the one in which the sole component of the 
gauge field is $r$-dependent instead of $y$-dependent: $A_0=a(r)$. What
we actually do, in this case, is that we consider a usual 4-dimensional
ansatz for the gauge field
configuration, that of a pure electric field, embedded in an extra
dimension. Then, the components of the bulk energy--momentum tensor
assume the form
\be
T^{t}_t=T^{r}_r=- \frac{e^{-4A}}{2}\,(\partial_r a)^2\,, \qquad 
T^\theta_\theta=T^\varphi_\varphi=T^5_5=
\frac{e^{-4A}}{2}\, (\partial_r a)^2\,.
\ee
The off-diagonal component of Einstein's equations can be integrated once 
again, leading to the same solution (\ref{ans1}) for $U(r,y)$. Turning next
to the diagonal components, we consider the sum of Eqs. (\ref{th-th}) and
(\ref{55}), which reduces to
\be
\frac{U-1}{e^{2A}\,r^2} + 
\frac{3}{e^{2A}\,r} \frac{\partial U}{\partial r} + 
\frac{1}{e^{2A}}\,\frac{\partial^2 U}{\partial r^2}= 
\kappa^2\,e^{-4A}\,(\partial_r a)^2 +
\biggl\{ \begin{tabular}{c} $-9 \Lambda  e^{-2A}$
\\ $0$ \end{tabular} \biggr \}\,,
\ee
or, alternatively,
\be
\frac{\lambda(y)}{r^4} + \sum_n \alpha_n\,r^{n-2} (n+1)^2=
\kappa^2\,e^{-2A}\,(\partial_r a)^2 +
\biggl\{ \begin{tabular}{c} $-9 \Lambda$
\\ $0$ \end{tabular} \biggr \}\,.
\ee
The above constraint is satisfied by the {\textit{solution}}
\be
\lambda(y) = \lambda_0\,e^{-2A(y)}\,, \qquad
(\partial_r a)^2=\frac{\lambda_0}{\kappa^2 \,r^4}\,, 
\qquad \biggl\{ \begin{tabular}{c}
$n=-1,\,a_{-1}={\rm const.} \quad {\rm and} \quad n=2,\,a_2=-\Lambda $ \\[2mm]
$n=-1,\,a_{-1}={\rm const.}$ \end{tabular} \biggr \}\,,
\ee
which is not admissible because of the discontinuous $\lambda$. It is
interesting to note that this ``would-be" solution restores the ``mass"
term $n=-1$ in the expression of the metric function, and/or
the cosmological constant term $n=2$, and
leads to the conventional expression for the ``electric" gauge
field, i.e. $a(r)=Q_e/r$, with $Q_e=\pm \sqrt{\lambda_0}/\kappa$.
However, the rearrangement of Eqs. (\ref{tt}, \ref{rr}) and
(\ref{th-th}, \ref{55}) leads to constraints similar to Eq. (\ref{con12}),
the only difference being the replacement of $2 V(\phi)$ by 
$e^{-4A}\,(\partial_r a)^2$. The combination of these two constraints 
can be rewritten as
\be
\frac{3U'^2}{2U^2} = -\kappa^2\,e^{-4A}\,(\partial_r a)^2 <0\,.
\ee
It is impossible to satisfy this constraint and, therefore, we are forced to
exclude the specific gauge field configuration used in the above analysis.

We will now address the most general case of having a five-dimensional
gauge field that depends both on bulk and brane coordinates. In order to
write down a sufficiently general ansatz for the gauge field, we turn for an
instant to the four-dimensional case and to the most general black hole
solution, i.e. the Kerr--Newman solution. The gauge field configuration
that gives rise to such a solution has the form~\cite{carter}
\be
A_\mu \, dx^\mu=\frac{Q_e r}{\rho^2}\,[\,dt -A_J\,\sin^2\theta\,d\varphi\,]
+ \frac{Q_m \cos\theta}{\rho^2}\,[\,A_J\,dt -(r^2+A_J^2)\,d\varphi\,]\,,
\ee
where $A_J$ is the angular momentum and $\rho^2=r^2 + A_J^2\,\cos^2\theta$.
Since here we are interested in black hole solutions that are characterized by
a spherical symmetry, we may set $A_J=0$, which reduces the black hole solution
to the Reissner--Nordstrom one. Note that the presence of the radial magnetic
field does not change at all the form of the solution for the metric
tensor: it merely replaces the $Q_e^2$ metric parameter by $Q^2=Q_e^2+Q_m^2$.
Therefore, for simplicity, we may ignore the $\theta$-dependence of the
$A_3$ component of the gauge field that gives rise to the radial magnetic
field. Coming back to the 5D case and generalizing the form of the gauge
field, we may write the following ansatz
\be
A_0=A_0(r,y)\,, \quad A_1=A_1(y)\,, \quad A_2=A_2(y)\,, \quad
A_3=A_3(y)\,, \quad A_5=A_5(r,y)\,.
\ee
Another simplification is in order: the spherical symmetry, which characterizes
our ansatz for the metric tensor (\ref{metric}), leads to the relation
$G^\theta_\theta=G^\varphi_\varphi$ for the Einstein tensor, which subsequently 
demands the same relation for the corresponding components of the bulk
energy--momentum tensor. It is straightforward to see that this leads to
the constraint $A_2=A_3=0$. That leaves us with the following expressions
for the components of $T^{M}_{N}$
\be
T^{t}_t=F_{tr}\,F^{tr} + F_{t5}\,F^{t5} -\frac{1}{4}\,F^2\,, \qquad
T^{r}_r=F_{tr}\,F^{tr} + F_{r5}\,F^{r5} -\frac{1}{4}\,F^2\,, \qquad
\ee
\be 
T^{\theta}_\theta=T^\varphi_\varphi= -\frac{1}{4}\,F^2\,, \qquad
T^{5}_5= F_{t5}\,F^{t5} + F_{r5}\,F^{r5} -\frac{1}{4}\,F^2\,, \qquad
T_{r5}=g^{tt} F_{tr}\,F_{t5}\,,
\ee
where
\be
F^2=F_{AB}\,F^{AB}=2 \Bigl(F_{tr}\,F^{tr} + F_{t5}\,F^{t5}
+ F_{r5}\,F^{r5}\Bigr)\,.
\ee

We first turn to the off-diagonal component (\ref{r5}): if we use
again the generalized ansatz (\ref{ans2}), this can be rewritten as
\be
\frac{\partial (r^2 U')}{\partial r}= -\sum_n \alpha_n'(y)\,(n+2)\,r^{n+1}=
2 \kappa^2\,e^{-2A}\,(\partial_r A_0)\,A_0'\,.
\label{off}
\ee
According to the above, and in agreement with the conclusions drawn in the
first two cases considered, when the $A_0$ component of the gauge field
depends only on one of the two coordinates, the only $y$-dependent term
that survives in the series of Eq. (\ref{ans2}) is the $n=-2$ term, and 
even this one turns out to be trivial according to the aforementioned
arguments. It is only in the case where this component has a non-trivial
dependence on both coordinates that other $y$-dependent terms, including
a ``mass" term ($n=-1$), might arise. Note that this is qualitatively
different from what happens in four dimensions:
there, an ``empty" spacetime around a spherical body
gives rise to a term proportional to $1/r$, while the introduction of a
gauge field leads to a higher term of ${\cal O}(1/r^2)$.

We now consider the sum of the pairs of Eqs. (\ref{tt}, \ref{rr}) and
(\ref{th-th}, \ref{55}). They have the form
\ba
&~& \frac{2(U-1)}{e^{2A}\,r^2} + 
\frac{2}{e^{2A}\,r} \frac{\partial U}{\partial r} +
\frac{U'^2}{2U^2} = -\kappa^2\,e^{-4A}\,(\partial_r A_0)^2 +
\biggl\{ \begin{tabular}{c} $-6 \Lambda  e^{-2A}$
\\ $0$ \end{tabular} \biggr \}\,,
\label{cona}\\[5mm]
&~& \frac{U-1}{e^{2A}\,r^2} + 
\frac{3}{e^{2A}\,r} \frac{\partial U}{\partial r} + 
\frac{1}{e^{2A}}\,\frac{\partial^2 U}{\partial r^2}= 
\kappa^2\,e^{-4A}\,(\partial_r A_0)^2 +
\biggl\{ \begin{tabular}{c} $-9 \Lambda  e^{-2A}$
\\ $0$ \end{tabular} \biggr \}\,,
\label{conb}
\ea
respectively. Taking the sum of the above two
constraints, (\ref{cona}) and (\ref{conb}), we arrive at the equation
\be
\frac{3(U-1)}{e^{2A}\,r^2} + 
\frac{5}{e^{2A}\,r} \frac{\partial U}{\partial r} + 
\frac{1}{e^{2A}}\,\frac{\partial^2 U}{\partial r^2}=
-\frac{U'^2}{2U^2} +
\biggl\{ \begin{tabular}{c} $-15 \Lambda  e^{-2A}$
\\ $0$ \end{tabular} \biggr \}\,.
\ee
or, alternatively, at
\be
\Biggl(\sum_n \alpha_n(y)\,(n+1)\,(n+3)\,r^{n-2}-
\biggl\{ \begin{tabular}{c} $15 \Lambda$
\\ $0$ \end{tabular} \biggr \} \Biggr)\,
\Bigl[1-\sum_m \,\alpha_m(y)\,r^m \Bigr]^2=
\frac{e^{2A}}{2}\,\sum_{n,m} \alpha'_n(y)\,\alpha'_m(y)\,r^{n+m}\,.
\ee
Collecting terms of the same order in the above expression, we may easily
see that severe constraints arise for the coefficients $\alpha_n$ and their
derivatives. For example, in the case where $\Lambda=0$, considering terms
up to $n=-3$ in the above series, we find that we always have $\alpha'_{-1}=0$
and either $\alpha_{-2}'=\alpha_{-3}'=0$ or even $\alpha_{-2}=\alpha_{-3}=0$.
In either case, it is clear that the assumed dependence on the bulk
coordinate of the 4D metric function $U$ is forbidden. For $\Lambda
\neq 0$, and for terms up to $|n|=2$, we arrive at the unique solution
$\alpha'_{-1}=0$, $\alpha_{-2}=\alpha_1=0$ and $\alpha_2=\Lambda$,
which is nothing else than the $y$-independent AdS/dS--Schwarzschild
metric function (\ref{4DAdS}). 
We may therefore conclude that even in the presence of a bulk gauge
field, which depends on both brane and bulk coordinates, black
hole solutions of the form (\ref{metric}), with the 4D metric function
given by the generalized expression (\ref{ans2}), are not allowed.

\sect{AN ALTERNATIVE APPROACH}

From the results drawn in the previous section, it became clear that
in the case of an empty bulk, apart from the contribution coming from
the five-dimensional cosmological constant, and/or in the presence of
conventional bulk matter in the form of either a five-dimensional
scalar or gauge field, black hole solutions that respect the
localization of gravity \`a la Randall--Sundrum and, at the same time,
can accommodate a non-trivial $y$-dependence of the four-dimensional
metric function cannot exist. The question that arises next is: if
solutions with $y$-dependence do exist, such that the value of the
horizon itself scales with the extra dimension, reaching a zero value
before the AdS horizon is reached, what is the nature of the bulk matter
that can support these solutions? And, moreover, what is the distribution
of bulk matter along the extra dimension? Does it mimic the localization
of graviton zero mode and black hole horizon and have its peak near the
3-brane, or does it spread all the way up to the AdS horizon?

In order to answer the above questions, we will follow an alternative
approach. We will demand that the 4D metric function $U$ has indeed
a non-trivial dependence on the bulk coordinate, which is consistent
with the desired behaviour at small and large distances from the brane,
and we will subsequently determine the components of the unknown bulk
energy--momentum tensor. Their profile in both the bulk and brane
coordinates will be checked and an approximate equation of state,
as well as some characteristic features of the bulk matter distribution,
will be determined. 

To this end, we consider the following ansatz for the line--element
of the 5-dimensional spacetime that serves the above purpose:
\be
ds^2=e^{2 A(y)}\,\biggl\{-\Bigl(1-\frac{w(y)}{r}\Bigr)\,dt^2 +
\Bigl(1-\frac{w(y)}{r}\Bigr)^{-1} dr^2
+ r^2(d\theta^2+ \sin^2\theta\,d\varphi^2)\biggr\} + dy^2\,.
\label{metric1}
\ee
The above ansatz corresponds to a generalized $y$-dependent form of the
RS--Schwarzschild solution (\ref{bstring}) in which the value of the
horizon of the four-dimensional black hole scales with the extra
coordinate. We will also assume that the conformal factor that multiplies
the 4D line--element has the same profile as in the RS model, 
$e^{A(y)}=e^{-\lambda |y|}$. The arbitrary function $w(y)$ will be
chosen to have the following behaviour :
$\lim_{y\,\rightarrow\,0}\,w(y)\,\rightarrow\,r_h$ and 
$\lim_{y\,\rightarrow\,\infty}\,w(y)\,\rightarrow\,0$.
In this way, we ensure that the induced metric on the brane is of the
Schwarzschild type and that the value of the black hole horizon vanishes
before the AdS horizon is reached.  

For the above spacetime background (\ref{metric1}), Einstein's equations
in the bulk take the form
\ba
3\,(A''+ 2 A'^2) +  \frac{2 A' w'}{(r-w)} + \frac{w''}{2(r-w)} +
\frac{3 w'^2}{4(r-w)^2} &=& \kappa^2\,\,[-\Lambda_B + T^{t}_t] \\[3mm]
3\,(A''+ 2 A'^2) -  \frac{2 A' w'}{(r-w)} - \frac{w''}{2(r-w)} -
\frac{w'^2}{4(r-w)^2} &=& \kappa^2\,\,[-\Lambda_B + T^{r}_r] \\[3mm]
 3\,(A''+ 2 A'^2) + \frac{w'^2}{4(r-w)^2} &=& \kappa^2\,\,
 [-\Lambda_B + T^{\theta}_\theta] \\[3mm]
6 A'^2 - \frac{w'^2}{4(r-w)^2} &=& \kappa^2\,\,
[-\Lambda_B + T^{5}_5]\\[3mm]
\frac{w'}{2r\,(r-w)} &=& \kappa^2\,T_{r5}\,.
\ea
If we adopt the RS choice for the warp factor, the terms proportional to $A'^2$ 
and $A''$ cancel exactly the bulk cosmological constant  and the brane term,
leaving the remaining terms involving derivatives of the unknown
function $w(y)$ to determine the components of the bulk energy--momentum
tensor $T^M_N$. It is straightforward to check that the resulting components
trivially satisfy the equation for the conservation of energy and momentum,
$D_M T^M_N=0$, for arbitrary $w$. Given the fact that the ansatz for
the four-dimensional metric function $U(r,y)$ considered above is a special
case of the more general ansatz (\ref{ans2}), it becomes clear that the
above non-constant, bulk matter distribution can never be attributed
to either a scalar or a gauge 5D field minimally coupled to gravity.

In order to investigate the properties of the aforementioned bulk matter
distribution, we will check its profile along both brane ($r$) and bulk ($y$)
coordinates. For this reason, we need to specify the metric function $w(y)$,
which apparently determines both the profile of the black hole horizon along
the extra dimension and the distribution of energy in the bulk. A choice that
respects the asymptotic conditions imposed on the $w$ function
both near and away from the brane is $w(y)=r_h\,e^{-a y^2}$, where $r_h$
is the four-dimensional value of the black hole horizon and $a$ an arbitrary,
positive constant. In this case, in the limit $y \rightarrow 0$, we
recover the Schwarzschild metric of a black hole located on the brane,
as demanded. On the other hand, in the limit $y \rightarrow \pm \infty$,
i.e. at large distances from the brane, the horizon of the black hole
shrinks to zero. Moreover, this particular choice is consistent with
the demand that the four-dimensional metric function $U(r,y)$ be
continuous along the extra dimension, since
$[U']_{y=0} \sim y\,e^{-a y^2}|_{y=0}= 0$, 
and the physical demand  that the bulk matter
distribution decay faster than the  AdS horizon.

For the aforementioned choice of the $w$-function, the components of the
bulk $T^M_N$ come out to be:
\be
\kappa^2\,T^t_t  =   \frac{r_h (4\lambda |y|+2a y^2-1)\,a e^{-ay^2}}{(r-w)}
+\frac{3 a^2 y^2 r_h^2 e^{-2ay^2}}{(r-w)^2} \,, \quad 
\kappa^2\,T^5_5 = -\kappa^2\,T^\theta_\theta 
=\frac{\kappa^2}{2}\,(T^t_t + T^r_r)\,,
\ee
\be
\kappa^2\,T^r_r=  -\frac{r_h (4\lambda |y|+2a y^2-1)\,a e^{-ay^2}}{(r-w)}
-\frac{a^2 y^2 r_h^2 e^{-2ay^2}}{(r-w)^2}\,, \quad 
\kappa^2\,T_{r5} = -\frac{a y r_h e^{-ay^2}}{r\,(r-w)}\,.
\ee
The behaviour of the three independent components of the energy--momentum
tensor in the bulk, $T^{t}_t$, $T^{r}_r$ and $T_{r5}$, with respect to the
$(r, y)$-coordinates, is depicted in Fig. 1. The figure shows an example
of a distribution of bulk matter capable of sustaining a localized black hole
solution. In terms of the $r$-coordinate, and for each AdS slice $y$=const.,
all the energy--momentum tensor components decrease as the distance from
the black hole horizon increases, and eventually vanish as we approach the
asymptotically flat regime $r \rightarrow \infty$. Near the black hole
horizon all components diverge, the diagonal ones faster, the off-diagonal
slower. The singularity of the energy--momentum tensor at the horizon 
is a reflection of an analogous singularity of the scalar curvature
($R \sim w'^2/(r-w)^2$). As we will shortly see, a more complicated
choice of metric might not have this problem. Since the singularity at
the horizon does not modify the issues of interest, namely the
localized nature of solutions and the curvature singularity structure at
infinity, we chose to stick with this simple choice and proceed.
%
%
%
\begin{center}
\begin{figure}[t]
\vspace*{2mm}
\begin{tabular}{c}
\epsfig{file=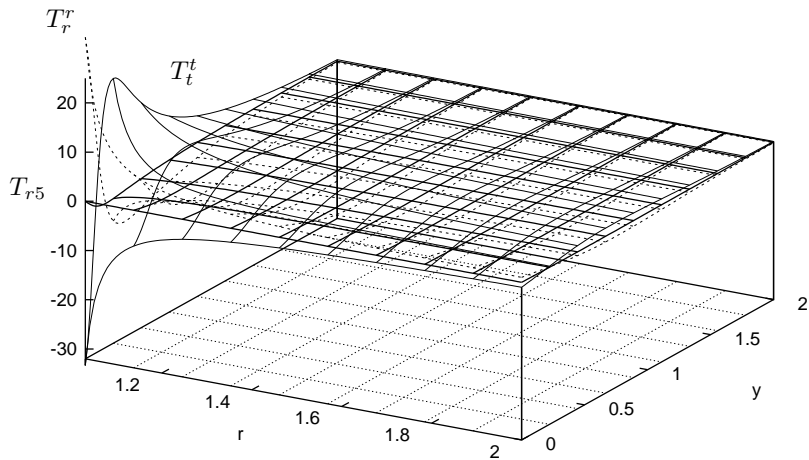, height=6cm}
\end{tabular}\\[10mm]
\caption{The plot depicts the profile of the three components of the
bulk energy--momentum tensor, $T^t_t$, $T^r_r$ and
$T_{r5}$, along the radial ($r$) and bulk ($y$) coordinate,
for $\lambda=a=r_h=1$ and $w(y)=r_h\,e^{-ay^2}$. 
Both profiles ``decay" very fast ensuring that all components
vanish at large distance from the brane and/or the horizon of the
black hole.}
\end{figure}
\end{center}

Turning to the $y$-profile of the bulk
matter and keeping $r$ fixed, we may easily see that the distribution
of matter is localized near, but off, the 3-brane. All components
rapidly decrease as one moves away from the brane, acquiring vanishing
values as $y \rightarrow \infty$, while, at the location of the brane,
all components are finite. The profile along the extra
coordinate becomes clearer in Fig. 2: the components $T^t_t$ and
$T^r_r$ are the only ones that have non-vanishing values at the
location of the brane, while all the remaining ones vanish. The distribution
of bulk matter is symmetric with respect to the brane. All
components reach their peak, in absolute values, slightly off, and on
both sides of, the brane thus creating a symmetric
``shell" that engulfs the brane.

\begin{center}
\begin{figure}[t]
\vspace*{2mm}
\begin{tabular}{c}
\epsfig{file=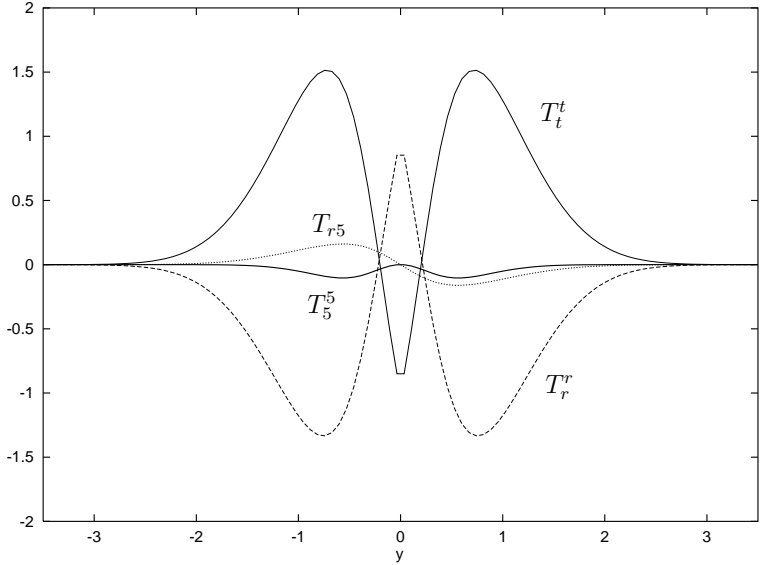, height=7.5cm}
\end{tabular}\\[10mm]
\caption{The profile of the bulk energy--momentum tensor components, in
terms of only the $y$-coordinate and for fixed $r=2 r_h$, is shown for
$\lambda=a=r_h=1$ and $w(y)=r_h\,e^{-ay^2}$. All components remain localized
near the brane, creating a symmetric ``shell".}
\end{figure}
\end{center}

If we define a bulk energy density and pressure in the following way:
$\rho=-T^{t}_t$, $p_i=T^{i}_i$, $p_5=T^{5}_5$ and $p_{r5}=T_{r5}$,
we may notice some more features, in the distribution of bulk matter,
emerging from Fig. 2. First, the dominant components $T^t_t$ and
$T^r_r$ behave, throughout the extra dimension, in such a way that
they always satisfy a {\it stiff} equation of state, i.e. $\rho=p_r$.
All the remaining components acquire a, relatively small, maximum value
slightly off the brane while they decay very fast as one moves close
to or away from the brane. Coming back to the dominant components,
we further observe that, although at the location of the brane we have
a physically acceptable situation with $\rho>0$ and $\rho+p_r>0$, both
components, $\rho$ and $p_r$, change sign at a particular point along
the extra dimension, thus leading to $y$-regimes where both the weak and
strong energy conditions are violated. Therefore, the result drawn
in the previous section that this energy distribution cannot be realized
by a conventional form of energy comes as no surprise. 

It is nevertheless important to check the behaviour of any curvature
invariant quantity that can be constructed in terms of the metric ansatz
(\ref{metric1}). The ``decaying'' value of the black hole horizon, as
one moves away from the brane, is ensured; however, one needs also to
investigate whether the black hole singularity is also ``localized''
near the brane. For this reason, we construct the $R_{MNRS} R^{MNRS}$
curvature invariant, which, for the RS choice of warp factor, is given
by the following expression
\ba
R_{MNRS} R^{MNRS} &=& 40 \lambda^4 + \frac{12 w^2 e^{4 \lambda |y|}}
{r^6} + 8 A' w'\biggl[\frac{w'^2}{(r-w)^3} + \frac{w''}{(r-w)^2}
\biggr] + \frac{2( w''^2 +7 \lambda^2 w'^2)}{(r-w)^2}\nonumber \\[2mm]
&+& \frac{w'^2}{(r-w)^3}\,\biggl[4 w'' + \frac{11 w'^2}{4 (r-w)}\biggr]
+\frac{2 w'^2 (3r-2w) e^{2\lambda |y|}}{r^3 (r-w)^2}\,.
\ea
The first term in the above expression is due to the presence of the bulk
cosmological constant and is present also in the case of a $y$-independent
four-dimensional metric tensor. The second term is the one that controls
the behaviour of the black hole singularity in the case of the
RS--Schwarzschild solution, Eq. (\ref{riemann}). We observe that the
profile of this term along the extra dimension, in this particular
case, is radically different. The presence of the additional scaling factor
$e^{-2 a y^2}$, which comes from the expression of the ``decaying'' black hole
horizon, causes this term to vanish at a large distance from the brane.
In this way, the singularity at the AdS horizon disappears and, moreover,
the black hole singularity itself ``decays'' as one moves away from
the brane, thus leading to the localization of the black hole.
However, there appear additional terms, which owe their existence
to the $y$-dependence of the four-dimensional metric tensor. These terms
give rise to a new singularity in the five-dimensional spacetime, which is
located near the black hole horizon and dies out as one moves away from
the brane. It can be easily shown that this new singularity is 
characteristic only of the particular five-dimensional spacetime which is
described by the metric tensor, Eq. (\ref{metric1}). Clearly, there are
more that one five-dimensional spacetimes, each one possibly with a different
global structure, that reduce to the same black hole solution
on the brane. An alternative choice of the five-dimensional 
spacetime, which also induces a Schwarzschild solution
on the brane but without possessing a horizon, might not have this problem.
As such, we consider the following ansatz
\be
ds^2=e^{2 A(y)}\,\biggl\{\frac{32 M(y)^3 e^{-r/2M(y)}}{r}\,
(-dv^2 + du^2)
+ r^2(d\theta^2+ \sin^2\theta\,d\varphi^2)\biggr\} + dy^2\,,
\label{metric1b}
\ee
which is a five-dimensional analogue of the Kruskal--Szekeres black hole
solution, $r$ being a function of ($v, u, y)$. Note that, on the brane,
both ans\"atze, (\ref{metric1}) and (\ref{metric1b}), reduce to the same
black hole solution: the two four-dimensional line--elements
are related by a coordinate transformation, $ (t,r) \leftrightarrow (v,u)$,
and describe the same 4D spacetime. However, the
embedding of these line-elements in an extra dimension breaks their
equivalence, a fact which is reflected in the different sets of curvature
invariant quantities determined for the 5D line-elements (\ref{metric1})
and (\ref{metric1b}). The above metric ansatz is not plagued by the
presence of a horizon and, in turn, no new singularity appears
either in the expression of the scalar curvature or in the components
of the bulk energy--momentum tensor. 
Unfortunately, for this alternative choice,
a decaying value of the black hole horizon fails to either eliminate 
the singularity at the AdS horizon or to localize the black hole
singularity. We might therefore conclude that the optimum choice of
the 5-dimensional metric tensor, which would be capable of sustaining
localized black holes without the appearance of new singularities,
is, in principle, a possible but rather difficult task.

Let us note, at this point, that the distribution of bulk matter around the
brane and the black hole horizon should not be allowed to destroy the
assumed staticity of the solution by generating an effective cosmological
constant on the brane. Going back to the action (\ref{action1}) of our
five-dimensional gravitational theory, substituting the 5D scalar
curvature in terms of the metric tensor (\ref{metric1}) and integrating over
the extra dimension, we can easily evaluate the expression for the
four-dimensional effective cosmological constant to be 
\be
\Lambda = \frac{1}{3}\,\Bigl(\sigma-
\frac{6\lambda}{\kappa^2}\Bigr) + \int\,dy\,e^{-4\lambda |y|}\,
\Bigl[\frac{w'^2}{4 \kappa^2 (r-w)^2} + {\cal L}_B\Bigr]\,.
\label{action4}
\ee
The first term in the above expression vanishes identically because of
the fine-tuning between the bulk cosmological constant and the brane
tension, which is valid in the original RS model and holds also in our
case. The second term comes from the extra $y$-dependence of the metric
due to the function $w(y)$ and the assumed bulk energy distribution that
supports this extra dependence. If we want $\Lambda$ to be zero,
then, the second term should also vanish, thus leading to a
constraint on the bulk Lagrangian. Note that, under this constraint,
no trace of the bulk matter distribution is left in the four-dimensional
effective theory and, therefore, any particular features of $T^M_N$
cannot possibly affect the four-dimensional system.

As shown above, the ansatz for $U(r,y)$ used in (\ref{metric1})
is not the only
choice that fullfills our constraints for the four-dimensional metric
function. Moreover, the ansatz (\ref{ans2}) used in the previous sections
is not the most general one, since the $(r, y$)-dependent part of $U(r,y)$
need not be in a factorized form. An alternative metric ansatz that also
respects our demands for the behaviour of the 4D metric function near
as well as far away from the brane is
\be
ds^2=e^{2 A(y)}\,\biggl\{-\Bigl(1-\frac{r_h}{\sqrt{r^2+y^2}}\Bigr)\,dt^2 +
\Bigl(1-\frac{r_h}{\sqrt{r^2+y^2}}\Bigr)^{-1} dr^2
+ r^2(d\theta^2+ \sin^2\theta\,d\varphi^2)\biggr\} + dy^2\,.
\label{metric2}
\ee
The above ansatz also reduces to the Schwarzschild solution in the limit
$y \rightarrow 0$, recovering the black hole metric function on the brane.
In the other asymptotic regime, $y \rightarrow \pm\infty$, the term
proportional to $r_h$ shrinks to zero, as demanded, for any value of $r$.
Following the same method as for the previous metric ansatz (\ref{metric1})
and isolating those terms in Einstein's equations whose presence is
attributed to the bulk matter distribution, we determine the components
of $T^M_N$ whose behaviour with respect to the $(r,y)$-coordinates is
shown in Fig. 3. From the figure it becomes evident that the particular
matter distribution is highly unphysical: although the behaviour of all
components near the brane is identical to the one of the previous case,
the situation changes radically at the other asymptotic regime of the
extra dimension. All components, apart from the off-diagonal one,
$T_{r5}$, that tends to zero, start to diverge rapidly, eventually
acquiring infinite values near the AdS horizon. This behaviour obviously
violates any desired notion of localization of matter around the brane, as
it is physically unreasonable to accept a diverging distribution of matter
at infinity in order to support a localized black hole solution near the
brane. We therefore conclude that a well-defined behaviour of the metric
function, near or away from the brane, does not guarantee the same
behaviour for the distribution of bulk matter.


\begin{center}
\begin{figure}[t]
\begin{tabular}{c}
\epsfig{file=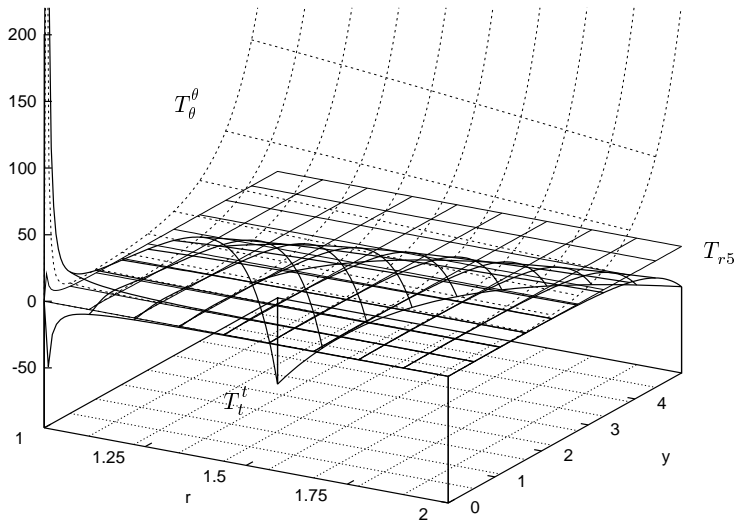, height=8cm}
\end{tabular}
\caption{The plot depicts the profile of the $T^t_t$, $T^\theta_\theta$
and $T_{r5}$ components of the bulk energy--momentum tensor, for the
metric ansatz (\ref{metric2}) and $\lambda=a=r_h=1$. All components (apart
from $T_{r5}$) diverge at $y$-infinity, violating any notion of localization
of matter around the brane.}
\end{figure}
\end{center}

\sect{CONCLUSIONS}

In this paper, we have addressed the issue of the localization of 
brane-world black holes along the extra dimension by investigating
the existence of $y$-dependent four-dimensional line--elements that
would allow for an extra scaling -- decaying -- of the physical value
of the black hole horizon in terms of the extra coordinate, while
reducing to a conventional black hole solution on the brane. 

Throughout the paper, a spherically symmetric, non-factorizable
$y$-dependent ansatz for the four-dimensional metric tensor was made,
i.e. $g_{\mu\nu}=e^{2A(y)}\hat{g}_{\mu\nu}(x,y)$, with the assumption
that, in the limit $y \rightarrow 0$, our ansatz should reduce to a 
well-known four-dimensional black hole solution. Apart from the desired
behaviour near as well as away from the brane, our ansatz was chosen to
satisfy the jump conditions with a continuous $\hat{g}_{\mu\nu}(x,y)$,
while the discontinuity at the brane was taken care of by
the warp factor. We restricted our investigation  by considering
the case $g_{tt}=-g_{rr}^{-1}$ and taking the warp factor to correspond
either to the simple Randall--Sundrum choice $A(y)=-\lambda |y|$ or to the
one of the RS--AdS/dS solutions. We believe that the essential features of this
analysis will remain true, even in the case that we generalize our ansatz
and move beyond these choices.

In the first part of the
paper, we considered the cases of empty bulk and bulk containing either
a scalar or a gauge field that might support such $y$-dependent solutions.
In particular cases, the most general form of the metric function $U(r,y)$
compatible with the equations of motion was derived by directly
integrating the off-diagonal component of Einstein's equations. In more 
complicated cases, a particular ansatz based on our assumptions and
physically motivated by the form of the previously derived solutions
was used. In all cases considered, of an empty, scalar-field- or
gauge-field-dominated bulk, no solutions supporting the desired $y$-dependence
of the 4D metric function and, thus, the scaling of the black hole horizon
with respect to the extra dimension, independently of the one
of the warp factor, were derived. The assumed $y$-dependence was 
inconsistent with either the continuity of the 4D metric function
$U(r,y)$ across the brane or the reality of the same function, and in
all cases the $y$-independent Schwarzschild or AdS/dS--Schwarzschild
solutions emerged as the unique solutions of the system. 

In the second part of the paper, we followed an alternative approach,
in which a particular ansatz that supports the scaling of the black hole
horizon with the extra dimension was given as input and the nature of the
corresponding bulk energy--momentum tensor was to be determined. The
ansatz described a $y$-dependent generalization of the Schwarzschild
solution with a Randall--Sundrum type of warp factor. The value of
the black hole horizon scaled as $r_h(y)=r_h e^{-a y^2}$, which
ensured the ``decaying", and eventually the vanishing, of the black
hole horizon as well as the continuity of the 4D metric function.
Knowing the exact form of the five-dimensional line--element, we
were able to solve for the components of the corresponding bulk
energy--momentum tensor and to check their behaviour, with respect to
the brane ($r$) and bulk ($y$) coordinates. According to our analysis,
for each AdS slice ($y=y_*$), the components of $T^M_N$ closely follow
the behaviour of the metric tensor by diverging near the black hole horizon
and vanishing asymptotically (as ${\cal O}(1/r)$) away from it. Regarding
their $y$-profile, all components turned out to respect the notion of
localization of matter around the brane: the matter distribution has
its peak slightly off the brane and rapidly decreases as we move away
from the brane. Both the bulk energy--momentum tensor and the black hole
horizon vanish well before the
AdS horizon is reached. The energy density and radial pressure of the
bulk matter are shown to satisfy a stiff equation of state throughout
the extra spacetime while the change of their sign in particular
$y$-regimes confirms the non-conventional type of nature of the bulk matter. 
The calculation of curvature invariant quantities, such as the square of the
Riemann tensor, reveals the disappearance of the singularity at the AdS
horizon and the localization of the black hole singularity around the brane.
However, a second, although localized, singularity appears at the black
hole horizon due to the singular behaviour of the bulk energy--momentum
tensor at this point. By considering an alternative choice for the
five-dimensional metric tensor, which describes a spacetime with a
distinctly different topology from the first one but which also reduces
to the same black hole solution on the brane, we were able to
demonstrate that this singularity is merely due to the particular choice
of the five-dimensional spacetime. Any additional choices, however,
considered for the five-dimensional spacetime, that were free from
the second singularity at the induced black hole horizon, were instead
plagued by the singularity at the AdS horizon. It would be therefore
interesting to investigate whether this bulk singularity, located either
at the AdS horizon or at a finite distance from the brane, is a generic
feature of every five-dimensional spacetime that induces a black hole
solution on the brane (as it is, for example, argued in the case of a
radiating brane on the grounds of the AdS/CFT correspondence \cite{AdS})
or whether an alternative, well-defined 5D spacetime indeed exists. We
leave this investigation for a future work.

As a concluding remark, let us note that our analysis cannot be considered
as a no-go theorem analysis for the existence of realistic localized black
holes by means of the decaying of their horizon value with the extra
dimension. The analysis in Secs. 3, 4 and 5 does not cover every possible
choice for the metric function $U(r,y)$. However, a general analysis along
the same lines, which would be able to exclude any form of $y$-dependent
metric function, is extremely difficult. In the same way, the ansatz
considered in Sec. 6 is only one of the many ans\"atze that might satisfy
our assumptions. Our analysis clearly suggests that the
localization of brane-world black holes strongly depends on the nature
of the bulk matter distribution. The localized black hole solutions, of
the type considered in this paper, were shown to demand an exotic form of
bulk matter. The quest for localized black hole solutions, supported by
a matter distribution that has a physically acceptable interpretation,
is to be continued.

\bigskip

{\bf Acknowledgements.} 
P.K. is deeply grateful to John March-Russell and Valery Rubakov for
useful discussions.
K.T. is thankful to A. Kehagias for a number of illuminating discussions.
P.K. and K.T. acknowledge the financial support of the EU RTN contract No.
HPRN-CT-2000-00148 and the EU RTN contract No. HPRN-CT-2000-00152.


\end{document}